# OODT: Obstacle Aware Opportunistic Data Transmission for Cognitive Radio Ad Hoc Networks

Xiaoxiong Zhong, *Member, IEEE*, Li Li, Yuanping Zhang, Bin Zhang, Weizhe Zhang, *senior Member, IEEE*, and Tingting Yang, *Member, IEEE*

*Abstract*—In recent years, a large number of smart devices will be connected in Internet of Things (IoT) using an ad hoc network, which needs more frequency spectra. The cognitive radio (CR) technology can improve spectrum utilization in an opportunistic communication manner for IoT, forming a promising paradigm known as cognitive radio ad hoc networks, CRAHNs. However, dynamic spectrum availability and mobile devices/persons make it difficult to develop an efficient data transmission scheme for CRAHNs under an obstacle environment. Opportunistic routing can leverage the broadcast nature of wireless channels to enhance network performance. Inspired by this, in this paper, we propose an <u>O</u>bstacle aware <u>O</u>pportunistic <u>D</u>ata <u>T</u>ransmission scheme (OODT) in CRAHNs from a computational geometry perspective, considering energy efficiency and social features. In the proposed scheme, we exploit a new routing metric, which is based on an obstacle avoiding algorithm using a polygon boundary 1-searcher technology, and an auction model for selecting forwarding candidates. In addition, we prove that the candidate selection problem is NP-hard and propose a heuristic algorithm for candidate selection. The simulation results show that the proposed scheme can achieve better performance than existing schemes.

## I. INTRODUCTION

WITH the development of computer networks, the Internet of Things (IoT) [1-3] has attracted much attention from researchers. As more smart devices access IoT, the increasing demand for spectral resources will result in overcrowding in Industrial Scientific Medical (ISM) bands. According to Federal Communications Commission (FCC), spectrum utilization measurements have indicated that many bands are typically underutilized over different spaces and times, resulting in considerable spectrum waste. IoT can benefit by using the promising Cognitive Radio (CR) [4] technology to enable more efficient spectrum utilization [5, 6]. However, dynamic spectrum availability and devices' heterogeneity and mobility make it difficult to develop an efficient data transmission scheme for cognitive radio ad hoc networks (CRAHNs). And then, the traditional pre-determined end-to-end routing protocols are not suitable for CRAHNs, which are difficult to maintain the routing table. Hence, how to design an efficient routing protocol in CRAHNs is challenging.

Opportunistic Routing (OR) [7-10] is a promising technology for improving routing performance through using the broadcast nature of wireless medium. In OR, a node broadcasts a packet to its neighbors, the node's neighbors then have the chance to receive/hear the packet, and hence, they can cooperate in packet forwarding. Because the next hops do not need to be determined in advance in OR, it is more suitable for CRAHNs. However, in CRAHNs, the traditional OR protocols have suffered from several limitations due to the dynamic nature of the available spectrum and primary user (PU), which play an important role in routing metric design and candidate forwarder selection. The papers [11-33] have proposed several different schemes. However, in physical environments, the transmission range of secondary user (SU) may contain various obstacles (walls or buildings), which brings to light several challenges in OR design. For example, in the campus, human carry mobile device (SU) to collect data and transmission data, which are deployed in a polygon area. The teaching buildings (obstacles) will affect the performance of data transmission in this environment. For improving network performance, we

This work was supported by the National Natural Science Foundation of China (Grant Nos. 61802221, 61802220, 61771086), the Natural Science Foundation of Guangxi Province under grant 2017GXNSFAA198192, the Key Research and Development Program for Guangdong Province 2019B010136001, and the Peng Cheng Laboratory Project of Guangdong Province PCL2018KP005 and PCL2018KP004. (*Corresponding authors*: Li Li and Tingting Yang).

Xiaoxiong Zhong is with Peng Cheng Laboratory, Shenzhen 518000, China, and also with Guangxi Key Laboratory of Trusted Software, Guilin University of Electronic Technology, Guilin 541004, China and the Graduate School at Shenzhen, Tsinghua University, Shenzhen 518055, China. (E-mail: xixzhong@gmail.com).

Li Li is with School of Computer Science, Shenzhen Institute & Information Technology, Shenzhen 518172, China and with the Graduate School at Shenzhen, Tsinghua University, Shenzhen 518055, China. (E-mail: lilihitcs@gmail.com).

Yuanping Zhang is with College of Computer and Educational Software, Guangzhou University, Guangzhou 510006, China. (E-mail: ypzhang12@gmail.com).

Bin Zhang is with Peng Cheng Laboratory, Shenzhen 518000, China. (E-mail: bin.zhang@pcl.ac.cn).

Weizhe Zhang is with the School of Computer Science and Technology, Harbin Institute of Technology, Harbin 150001, China, and with Peng Cheng Laboratory, Shenzhen 518000, China. (E-mail: wzzhang@hit.edu.cn).

Tingting Yang is with School of Electrical Engineering and Intelligentization, Dongguan University of Technology, Dongguan 523000, China. (E-mail: yangtingting820523@163.com).

should find an obstacle avoiding data transmission path by an efficient method. How can we guarantee data transmission efficiency for opportunistic transmission scheme with considering the existence of obstacles in CRAHNs? In fact, there have been several obstacle-aware routing protocols proposed recently [34-37]. However, they only addressed traditional routing design with obstacles. As far as we know, there is no previous work addressing obstacles in OR protocol over wireless networks. To fill the gap, our goal is to design an efficient opportunistic data transmission scheme with obstacles for CRAHNs.

In this paper, we propose a novel OR scheme based on an obstacle avoiding algorithm and an auction model for selecting forwarding candidates from a computational geometry perspective, which is named <u>O</u>bstacle aware <u>O</u>pportunistic <u>D</u>ata <u>T</u>ransmission, OODT, to achieve efficient data transmission in CRAHNs under obstacle environments. The contributions of this paper are listed as follows.

● We present a new obstacle avoiding algorithm from a computational geometry perspective. The algorithm is modeled as a polygon search by a boundary 1-searcher, which can easily find available candidates for data communication in an obstacle environment. In addition, we prove the condition for a given polygon $P$ is boundary 1-searchable and the efficiency of the proposed search algorithm.

● We prove that the candidate selection problem in the CRAHNs is NP-hard. Hence, we solve the problem by a heuristic algorithm. In the algorithm, a new routing metric is proposed, which jointly considers social features and energy efficiency. Based on this metric, we propose a novel candidate selection algorithm based on an auction model and derives the optimal bid price for this auction model, which is affected with the number of candidates and their social features, energy and expected transmission count (ETX).

● Through evaluation, we compare OODT protocol for CRAHNs with some existing works, which performs well in terms of packet delivery ratio (PDR), expected cost of routing, average end-to-end delay, and network lifetime.

The rest of this article is arranged as follows. Section II presents the related works. Section III describes the proposed scheme, OODT, which includes three key components: a system model, an obstacle avoiding algorithm, and candidate selection based on an auction model. Section VI evaluates the performance of OODT by simulations. Finally, Section V gives the conclusion of this paper.

## II. RELATED WORK

### A. Opportunistic data transmission in CRAHNs

Recently, some OR schemes have been proposed for CRAHNs. Pan *et al*. [11] exploited a new routing metric for OR, which considers packet delivery rate to prioritize and optimally select forwarding candidates. Badarneh *et al*. [12] proposed a novel OR protocol based on spectrum available time and data transmission times. Similarly, our previous work [13] exploited the feature of available channel to improve the performance of OR. Lin *et al*. [14] put forward to a spectrum aware OR for single-channel cognitive radio networks (CRNs), provisioning end-to-end QoS. Furthermore, Lin *et al*. [15] induced a novel OR for regular and large-scale multi-channel CRNs, which exploits spectrum map established on local sensing information. In the first scheme SMOR-1, it exploits relay selections based on link transmission qualities. For the SMOR-2, it induces stochastic geometry for geographic OR for cooperative diversity. Liu *et al*. [16] proposed a novel OR for CRNs based on heterogeneous channel occupancy patterns, exploiting the statistical channel usage and the physical capacity of wireless channels in routing decision. In order to select the best forwarding candidate set, Liu *et al*. [17] further studied the OR and presented an OR protocol in multi-channel CRNs. In [18], we proposed a novel OR scheme, CANCOR, for multi-channel CRNs, which incorporates channel assignment in OR protocol for multi-channel multi-radio CRNs. Zheng *et al*. [19] proposed a multi-channel spectrum aware OR for CRNs from a network coding perspective. They exploit the link delay for routing metric. How *et al*. [20] proposed an OR protocol in multi-hop CRNs, which is oriented for service differentiation, which is a cross-layer solution with considering spectrum availability and delay about spectrum switching and queuing. Tang *et al*. [21] presented a geographical OR for CRNs from a network coding perspective, CGOR. The goal of CGOR is that minimize the average number of transmissions. Barve *et al*. [22] proposed reinforcement learning based OR for mobile CRNs, which jointly consider channel assignment, MARL. In MARL, it provides an online manner for updating routing information, which can successfully explore transmission opportunities. Cai *et al*. [23] presented a new mathematically framework that minimizing the delay for the proposed OR in multi-channel CRNs form a cross-layer perspective, which includes channel sensing, relay selection and packet division. Dai *et al*. [24] presented a multi-layer forwarder set selection algorithm for OR, MOR, which includes main forwarder set and backup forwarder set. MOR can be adaptive to the situation that PU suddenly appears on the using channel. Lin *et al*. [25] proposed a novel statistical QoS control mechanism through cooperative relaying from a network coded OR perspective, realizing virtual MIMO communications at session level. Lin *et al*. [26] proposed a cognitive and opportunistic relay selection scheme for cognitive M2M networks, which can mitigate destructive interference and efficiently selection the forwarders.

### B. Social aware data transmission in CRAHNs

In recent years, some researchers have focused on social features for data transmission in CRAHNs. The social features of CRAHNs are the social relationships with other objects, autonomously with respect to humans, e.g., friendship, clustering coefficient, community. As we know that in this network, the node will have more chances to forwarding data packets to its neighbors that have the same interest or is in the same community, which is an important factor in routing design, enhancing data transmission. Wu *et al*. [27] proposed a new concept, PU community, to describe the diversity of PUs geographic distribution in a given period of time. Based on PU community, they presented a reliability-driven routing scheme to improve the end-to-end routing reliability. Jing *et al.* [28] proposed a social-aware opportunistic routing and relay selection scheme, SoRoute, which first predicts the link reliability based on a new social-relationship-aware mobility model and then fuses the relationships of SUs to make a routing and relay decision. In this scheme, different prediction schemes are employed for the nodes with different relationships. Ji *at al*. [29] proposed to study the joint routing and scheduling problem for CRAHNs by considering the social behaviors of PUs, which exploits a normal distribution to approximately describe the PU behaviors. Eryigit *at al*. [30] focused on the cooperator set selection problem in CRAHNs considering social ties, i.e. which SUs to ask for cooperation so that resulting throughput and sensing accuracy are maximized with detection and false alarm probability constraints. Similarly, in our previous work [31], we presented a social and trust aware OR in CRAHNs, which exploits social ties and trust for multi-flow CRAHNs. Hou *at al*. [32] studied the social properties, including community, friendship, and individual selfishness of cognitive radio networks and analyzed the effect of these social properties on the performance of routing protocols. Lu *et al.* [33]considered both the primary and secondary activities in the actual accessible whitespace from a social activity perspective, and proposed a greedy spectrum aware routing algorithm for CRAHNs.

*C. Obstacle aware data transmission in wireless networks*

However, none of the above works systematically combines obstacle detection and routing design to improve the performance of CRAHNs. In [34], Sun *et al*. proposed a pre-processed cross link detection protocol, for geographic routing in mobile ad hoc network, which extracts an almost planar subgraph from a realistic network graph, instead of a unit disk graph, for face routing and makes the greedy-face-greedy geographic routing work correctly in realistic environments with obstacles. Wu *et al*. [35] presented a voronoi-trajectory based hybrid routing for ad hoc networks from an obstacle perspective. In their scheme, they choose the shortest obstacle-avoiding path according to the rule: whether the destination node is in the source node's two hop range. Xie *et al*. [36] proposed a heuristic tour planning algorithm to find an obstacle avoiding shortest tour based on spanning graphs in wireless sensor networks which divides the whole network region into the same size grid cells. However, they only consider obstacles in the traditional routing design.

To the best of our knowledge, the traditional OR design works in CRAHNs and obstacle aware data transmission in traditional wireless networks have the following disadvantages:

1) Most of the existing candidate selection algorithms of OR in CRAHNs were proposed based on the environment without obstacles and they perform well only when the assumptions do hold. However, the data transmission in real-world CRAHNs with obstacles are often too complex to be designed by traditional OR. Till now, there is still little attention being paid on applying obstacle detection in candidate selection algorithm of OR, which will cause a poor network performance.

2) Due to the features of OR, the obstacle aware data transmission schemes designed by traditional routing protocols are not capable of depicting the data transmission for OR, which degrades the evaluation and the optimization on them.

In summary, the current research has addressed opportunistic data transmission in CRAHNs from different aspects. However, as discussed above, they do not systematically combine obstacle detection and OR to improve the performance of CRAHNs, and they only consider obstacles in the traditional routing design. Hence, how to design an efficient obstacle aware opportunistic data transmission scheme for CRAHNs is a challenging issue. This paper aims to propose a solution to address this problem.

### III. OBSTACLE AWARE OPPORTUNISTIC DATATRANSMISSION SCHEME FOR CRAHNS

In this section, the main idea of the proposed OR with obstacles for CRAHNs was presented, referred to as Obstacle aware Opportunistic Data Transmission, OODT, including the system model, obstacle avoiding algorithm and candidate decision (selection and prioritization).

*A. System model*

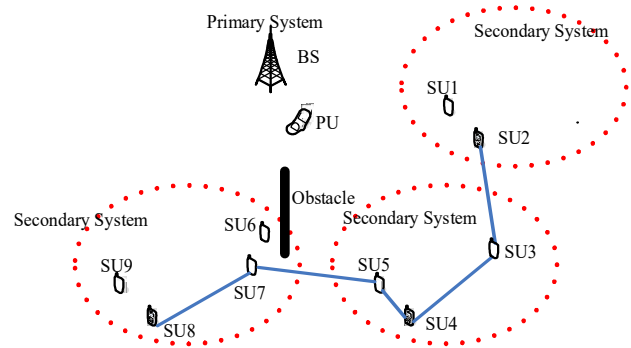

Fig. 1. Cognitive radio ad hoc networks (CRAHNs) with obstacle.

In OODT, an interweave model [18] will be used, that Secondary Users (SUs) can communicate among them only if no Primary User (PU) occupied the corresponding channel. We consider a time slotted multi-hop CRAHNs, in which node is equipped with $R$ radios working in a half-duplex model. Each SU has the ability for sensing channels and changing channel, which communicates between them with a fixed slot duration $T$, including a sensing period $T_s$ and a data transmission period $T_t$. The usage pattern of a given channel follows an independent ON/OFF state model with the lengths of busy and idle periods exponentially distributed with rate parameters $\lambda_{busy}$ and $\lambda_{idle}$ over busy and idle transition respectively. Note that the obstacles (walls or buildings) only affect SU communications in this paper. In SU's communication, it may be affected by the obstacles. As shown in Fig.1, the data transmission between SU2 and SU8 will select SU7 as its relay node, without selecting SU6 due to the obstacle. We also assume that each SU has the same fixed maximum transmit power over a given channel throughout the paper. In addition, a Rayleigh fading channel model is assumed to describe the fading channel between any two SUs. The average received signal power may be affected by shadowing from large obstacles, such as trees, buildings, or mountains. Measurements have shown that the path loss variation at a particular distance due to shadowing effect is a random variable with zero mean log-normal distribution. The shadowing is generally modelled as lognormal distribution [37, 38].

B. *Polygon search by a boundary 1-searcher*

In reality, there are some obstacles in CRAHNs, which can significantly affect SUs' data transmission. Hence, if we can detect the nodes in advance that are not sheltered from obstacles, then it is easier to determine the forwarding candidates. Next, we give an obstacle avoiding algorithm called a polygon search by a boundary 1-searcher.

First, we provide the preliminaries of the polygon search. A simple polygon $P$ consists of $n$ ($n \geq 3$) vertices and $n$ edges connecting adjacent vertices. The boundary of polygon $P$ is denoted $\partial P$, which contains all vertices and edges of $P$. Let $P'$ be the inner area of $P$. Two points $(a,b) \in P$ are considered visible to each other if and only if $\overline{ab} \subseteq P$, where $\overline{ab}$ denotes the line segment connecting $a$ and $b$ [39]. The vertices preceding and succeeding $v$ in clockwise order are denoted $Pred(v)$ and $Succ(v)$, respectively. A vertex of $P$ is reflex if its interior angle is strictly larger than 180°. Next, we give some symbol description. Suppose that $r$ is a reflex vertex. $Back(r)$ is the backward ray of $r$, which is the ray started from $Succ(r)$ to $r$ and the end point is the point on $\partial P$ where the ray leaves the polygon for the first time. Similarly, $Forw(r)$ is the forward ray of $r$, which is the ray started from $Prec(r)$ to $r$ and the end point is the point on $\partial P$ where the ray leaves the polygon for the first time. The line segments $\overline{rForw(r)}$ and $\overline{rBack(r)}$ are called the chords of the polygons as shown in Fig. 1.

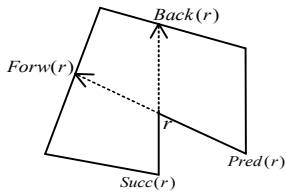

Fig. 2. Some concepts in a simple polygon.

Next, we introduce the *V-space*, *V-diagram* and the *skeleton V-diagram* [39]. We label the vertices $P$, 0, 1, 2, ..., $n$-1, clockwise around the boundary $\partial P$ for a given $n$-sided simple polygon $P$. We set an arbitrary boundary point (in our OR, it is a candidate) as the start point, and then calculate all distances along $\partial P$ clockwise from the start point. Let $D = |\partial P|$ denote the length of $\partial P$. For $x \in R$, $x$ is the point on $\partial P$ located at a distance $x - kD$ from the start point, where $k$ is an integer such that $0 \leq x - kD \leq D$. Thus, we also consider $x$ as a point on $\partial P$. Conversely, we often consider a point on $\partial P$ a real number as well representing that position.

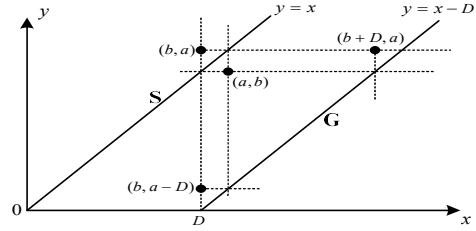

Fig. 3. Visibility space ($D = |\partial P|$).
Point ($b$, $a$-$D$) and ($b$+$D$, $a$) are symmetric to ($a$, $b$).

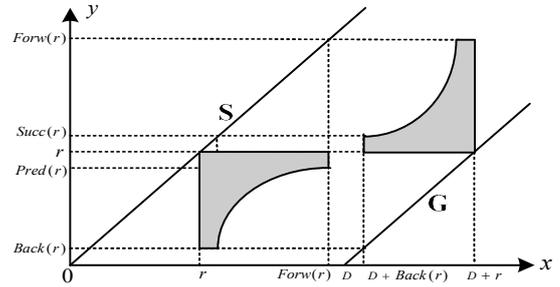

Fig. 4. SE and NW barriers due to reflex vertex $r$ ($D = |\partial P|$)

The visibility space [39] (V-space for short), denoted by $V$, consists of the infinite area between and including the lines $y = x$ ($x, y \in R$, the start line **S**) and $y = x - |\partial P|$ (the goal line **G**), as shown in Fig. 2, where $D = |\partial P|$. Therefore, we have $(x, y) \in V$ if and only if $x - |\partial P| \leq y \leq x$.

According to the visibility diagram [39] (V-diagram for short) for a given polygon, an idealized form is shown in Figure 3, and each gray shape is considered a *barrier* [39], i.e., an *obstacle* in this paper. A barrier whose corner is touching line **S** (or **G**) is called a southeast (or northwest) barrier (obstacle) of $r$, denoted as $SE(r)$ (or $NW(r)$) *barrier*. Figure 4 shows that reflex vertex $r$ gives rise to two barriers $SE(r)$ and $NW(r)$. $SE(r)$ have two special line segments, one is from ($r$, $r$) to ($r$, $Back(r)$), called the *southward line* of $r$, denoted by $sl(r)$, the other is from ($r$, $r$) to ($Forw(r)$, $r$), called the *eastward line* of $r$, denoted by $el(r)$. Symmetrically, two special line segments exist in $NW(r)$, one is from ($D + r$, $D+ r$) to ($D +r$, D+$Forw(r)$), called the *northward line* of $r$, denoted by $nl(r)$, the other is from ($D + r$, $D + r$) to ($D + Back(r)$, $r$), called the *westward line* of $r$, denoted by $wl(r)$. All these four line segments are called the *bones* of $r$.

Because the topology (bones) of the *V*-diagram is a part of the essential information about searchability in our proposed scheme, we can only extract the topological information from the *V*-diagram and eliminate other non-essential features. Therefore, by using two bone segments to replace a barrier in the *V*-diagram, the *skeleton V-diagram* is constructed. The construction details are provided in [39].

Hence, we can obtain characterizations of simple polygons to test whether a given polygon can be searched by a boundary 1-searcher and have the following results.

**Lemma 1** *A given polygon is boundary 1-searchable if and only if there is a search path in its pruned skeleton.*

**Theorem 1** *A given polygon P is not boundary 1-searchable if and only if one of the followings is true:*

*C1: Each boundary point $p$ is in the inner chain part of a one-side bi-tangent.*

*C2: Each boundary point $p$ in the inner chain part of a double left (or right) bi-tangent.*

*C3: For all the boundary points, some are in the inner chain part of one-side bi-tangents, and the others are in the inner chain part of double left (or right) bi-tangents.*

*C4: There exists a one-side bi-tangent constructed by u, v (say) and restricted reflex vertex w (say), which is different from u and v, in the inner chain part of this one-side bi-tangent such that w and u construct a left bi-tangent or w and v construct a right bi-tangent.*

**Proof**: see Appendix A for the proof.

**Theorem 2** *It takes $O(n)$ time and space to determine whether a given polygon is boundary 1-searchable.*

**Proof**: see Appendix B for the proof.

By the proof of **Theorem 2**, we can easily obtain the following result.

**Corollary 1** *It takes $O(n)$ time to determine all restricted reflex vertices in a given polygon.*

### C. Search strategy

The name of the algorithm is abbreviated to **BSA**, which is the basis of the obstacle aware candidate selection algorithm, and the implementation of the algorithm is as follows.

First, input a sequence of points of polygon $P$, then from step 2 to step 9 to test whether $P$ can be searched by a boundary 1-searcher. From step 10 we start to search by using our strategy. At the beginning, to determine whether $P$ has reflex vertices, if no, $P$ is a convex polygon. At that moment, at step 12 we select an arbitrary point $q$ as the start point of searching, and let the 1-searcher $s$ fix at the point $q$, the endpoint $f$ of the flashlight clockwise moves along $\partial P$ from $q$, when $s$ and $f$ overlap for the second time at the boundary point $q$, the search is finished, where the first overlap between $s$ and $f$ happens at the beginning of search. If there exist reflex vertices in $P$, then we find out an unrestricted reflex vertex as the start point to search, moreover, let $s$ fix at the current reflex vertex, and $f$ clockwise moves along $\partial P$ starting from the current point at step 16. In the following, we illustrate the implementation of **BSA** with the Figure 5 and the Figure 6.

Next, we detect whether $s$ and $f$ overlap for the second time at one boundary point during the searching at step 17. If the condition of step 17 is true, then, output a search schedule, the search is finished; if false, it must turn up the situation of step 20 or step 23. We take Figure 8 for an example to illustrate the processes of searching. We first find out an unrestricted reflex vertex as shown the vertex 4 in Figure 8, and let $s$ fix at vertex 4, let $f$ clockwise move along $\partial P$ starting from the vertex 4, till it encounters the reflex vertex 7. At the moment, if $f$ keeps on moving, resulting in invisible areas viewed from $s$, then let $f$ stay at the vertex 7, and $s$ counterclockwise moves along $\partial P$ to the next reflex vertex such as the vertex 1. If $s$ and $f$ are always visible when $s$ moves to the vertex 1, let $f$ stay at the vertex 7 until $s$ gets to the vertex 1. Later, go to step 16 and then to determine whether the condition of step 17 is true. If true, the search is complete and output a search schedule; if false, keep on running the following steps until $s$ and $f$ overlap for the second time at some boundary points of $P$, and output a search schedule.

In Figure 5, if $s$ and $f$ start search from the reflex vertex 9. First let $s$ fix at the vertex 9, and $f$ clockwise moves along $\partial P$ starting from the vertex 9. When $f$ gets to the reflex vertex 11, if $f$ keeps on moving, resulting in invisible areas viewed from $s$. Meanwhile, let $f$ stay at the vertex 11, and $s$ counterclockwise moves along $\partial P$ to the next reflex vertex such as the vertex 7. But in this process, as long as $s$ moves a very short distance from the vertex 9 to the vertex 8, $s$ and $f$ are not visible from each other. At the moment, we must adjust the position of $f$ in order to maintain the visibility of $s$ and $f$, that is, $f$ must jump from the vertex 11 to the vertex 9, till $s$ arrives at the next reflex vertex 7. Then, go to step 16 and to determine whether the condition of step 17 is true. If true, output a search schedule; if false, keep on running the following steps until $s$ and $f$ overlap for the second time at some boundary points of $P$, and output a search schedule.

As for the situation of step 23, we take Figure 6 for an example to illustrate the processes of searching. If $s$ and $f$ start search from the reflex vertex $a$, first let $s$ fix at the vertex $a$, then, let $f$ clockwise move along $\partial P$ to the point $c$. And if $f$ keeps on moving, it will jump to the reflex vertex $b$, resulting in the areas which have been cleared are recontaminated. At that moment, let $f$ stay at point $c$, and $s$ counterclockwise moves along $\partial P$ to the next reflex vertex such as the vertex $b$ in Figure 9. During the moving of $s$, it will also come across the situation of step 26 or step 28.

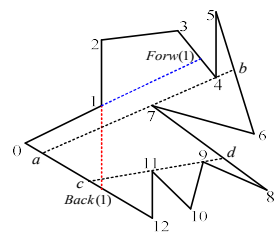

Fig. 5. An example polygon $P$.

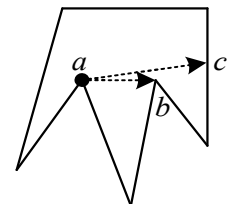

Fig. 6. An example polygon in which reflex vertex $a$ as a start point of searching.

**Algorithm 1** Boundary 1-searcher's Search Algorithm (**BSA**)

1: **Input:** a polygon $P$ as a sequence of points $(p_0, p_1, ..., p_{n-1})$
2: to determine whether $P$ is a LR-visible polygon
3: **if** $P$ is not a LR-visible polygon **then**
4: **goto 32**
5: **end if**
6: to determine whether $C_i$ ($i=1, ..., 4$) of **Theorem 1** is established
7: **if** one of $C_i$ is established **then**
8: **goto 32**
9: **end if**
10: to determine whether there exist reflex vertices in $P$
11: **if** no reflex vertex in $P$ **then**
12: take an arbitrary point $q$ as the start point, and let $s$ fix at $q$, $f$ clockwise moves along $\partial P$ from $q$, when $s$ and $f$ overlap for the second time at $q$, **goto 33**
13: **else**
14: select an unrestricted reflex vertex as the start point
15: **end if**
16: let $s$ and $f$ start search from the current reflex vertex, $s$ fixes at the current reflex vertex, and $f$ clockwise moves along $\partial P$. During searching, it must ensure that the clear portion of polygon is always to the left of flashlights $\overline{sf}$, as viewed from the 1-searcher.
17: **if** $s$ and $f$ overlap for the second time at one boundary point **then**
18: **goto 33**
19: **else**
20: **if** this case ($f$ moves along $\partial P$ to a reflex vertex $p$, and if $f$ keeps on moving, it will jump over a section of boundary chain and reach a boundary point $t$ which is the endpoint of $\overline{st}$ (there is an infinite small angle between the line $\overline{st}$ and the line $\overline{sp}$). This situation forms invisible areas on the left of flashlight $\overline{sf}$) occurs **then**
21: let $f$ stay at reflex vertex $p$, and $s$ counterclockwise moves to the next reflex vertex along $\partial P$
22: **end if**
23: **if** this case ($f$ moves along $\partial P$ to a non-reflex vertex $c$, and if $f$ keeps on moving, it will jump over a reflex $p'$, such that the areas which have been cleared are re-contaminated) occurs **then**
24: let $f$ stay at point $c$ and $s$ counterclockwise moves to the next reflex vertex along $\partial P$
25: **end if**
26: **if** $s$ and $f$ are always visible when $s$ moves to the next reflex vertex **then**
27: let $f$ stay at the current point until $s$ arrives at the next reflex vertex, **goto 16**
28: **else**
29: adjust the position of $f$, and make $s$ and $f$ always visible until $s$ arrives at the next reflex vertex, **goto 16**
30: **end if**
31: **end if**
32: **Output:** $P$ is not boundary 1-searchable
33: **Output:** a search schedule of $P$

**Theorem 3** *Using **BSA** to search a given boundary 1-searchable polygon will minimize the distance traveled by the searcher. Moreover, the maximum distance is less than $2|\partial P|$.*

**Proof:** From the construction and analysis of the **BSA** algorithm, in the best case, the searcher stays at the start point and will search an entire polygon completely in a round. Because $f$ is allowed to move at the beginning in **BSA**, as long as the condition of step 17 of **BSA** is established, $s$ does not need to move. In the worst case, the searcher only needs to travel along $\partial P$ less than twice when a polygon is searched completely. Moreover, because we select an unrestricted reflex vertex as a start point in **BSA**, moreover, the searcher $s$ keeps on moving forward along $\partial P$ counterclockwise, without any jump and back. When $s$ moves along $\partial P$ in a round and returns to the start point, some regions of $P$ may be recontaminated. The distance traveled by $s$ is less than two rounds, and the entire region must be clear, if $P$ is boundary 1-searchable. When $s$ moves two rounds along $\partial P$ and returns to the start point, the search processes are same as the previous.

In summary, the maximum distance traveled by searcher is less than $2|\partial P|$. Hence, the above theorem is established.

**Theorem 4** *Using **BSA** to search a given n vertices polygon P, a search schedule can be output in $O(m)$ time, if it exists, where $m$ ($< n^2$) is the number of search instructions reported.*

**Proof:** From step 2 to step 9 of **BAS**, we can see that to determine whether a given polygon is boundary1-searchable needs $O(n)$ time. The goal of the remain steps is to find a search schedule, if it exists. Base on the analysis of algorithm we can conclude that output a search schedule only needs $O(m)$ time, where $m$ ($< n^2$) is the number of search instructions reported. Now, we analyze $m$ (In fact, $m$ is just the total number of vertices traveled by $s$ and $f$.) in the worst case: the endpoint $f$ of the flashlight passes through $n/2$ vertices as long as $s$ reaches a vertex. At that moment, $m < n^2$ (when $s$ gets to a reflex vertex, $f$ passes through $n/2$ vertices, and $s$ moves from the current reflex vertex to the next one, $f$ also passes through $n/2$ vertices in order to maintain the visibility with $s$. By **Theorem 3**, the maximum distance traveled by $s$ is less than $2|\partial P|$, and the number of the vertices in $P$ is $n$), which is the upper bound. This completes the proof.

Through the boundary 1-search algorithm, we can calculate the number of nodes observed by a certain node. If the polygons (formed by the network topologies) are not boundary 1-searchable, we delete a point randomly and execute **Algorithm1**. Hence, the data transmission performance will significantly improve, considering nodes that are in the obstacle zone. Next, we discuss candidate decision based on the proposed obstacle aware data transmission algorithm.

*D. Auction based candidate selection*

In this section, the details of the data forwarding strategy will be presented, which includes obstacle aware algorithm and candidate selection algorithm.

*1) Social ties*

Social ties can be characterized according to historic information, e.g., the frequency, duration of contact and social features. Hence, the social ties of nodes $i$ and $j$ can be calculated:

$$ST_{i,j}(T) = \chi SPM_{i,j}(T) + (1-\chi)socsim_{i,j}(T) \quad (1)$$

where $SPM_{i,j}(T)$ is social pressure metric of nodes $i$ and $j$ [44] in the duration $T$, $\chi (\in [0,1])$ is a weight parameter, and $socsim_{i,j}(T)$ is the social similarity of nodes $i$ and $j$ in the duration $T$, which can be expressed:

$$socsim_{i,j}(T) = com_{i,j}(T)/(n_i(T)+n_j(T)) \quad (2)$$

where $com_{i,j}(T)$ is the number of common neighbors of nodes $i$ and $j$ exploiting our proposed obstacle avoiding scheme (**Algorithm1**) in the duration $T$, $n_i(T)/n_j(T)$ is the number of node $i$'s / $j$'s one-hop neighbors (which are boundary 1-searchable) in the duration $T$.

*2) Energy consumption model*

In our scheme, the energy consumption $E_{iC}(T)$ for successfully one packet transmission from node $i$ to its downstream neighbors in the duration $T$, which includes $E_{iF}(T)$ consumed to forward a packet, $E_{iR}(T)$ consumed to receive/hearing a packet, and $E_{iACK}(T)$ consumed to send an feedback packet. Thus, we can obtain

$$E_{iC}(T) = E_{iF}(T) + n_i(T) \times E_{iR}(T) + E_{iACK}(T) \quad (3)$$

In this scheme, we assume the energy function is subject to a (0, 1) uniform distribution.

*3) Two-layer auction model for candidate selection*

In proposed scheme, we classify the candidates into two types of candidate forwarding sets (*CFS*) *CFS₁* and *CFS₂*, where *CFS₁* is a primary set, and *CFS₂* is a backup set and is a set featuring a common node belonging to a different node's candidate set at the same time. In this auction model, the source node is the auctioneer and the nodes which are in candidate set are the bidders. The cost of the $i^{th}$ bidder is $v_i$, and the bidder's offered price of node $i$ is $b_i^p$ where $i$ is in *CFS₁* and $b_i^s$ is the bidder price of node $i$ where $i$ is in *CFS₂*. Note that we have $CFS_1 \cap CFS_2 = \varnothing$. Hence, we can obtain the node $i$'s payoff function:

$$u_i = \begin{cases} b_i^p - v_i, & i \in CFS_1 \\ b_i^s - v_i, & i \in CFS_2 \\ 0, & \text{otherwise} \end{cases} \quad (4)$$

In a given set *CFS* in the networks, select a subset *CFS′* such that $u_0(CFS')$ is maximized over all subsets, can be called as a candidate selection problem.

**Theorem 5** *The candidate selection problem is NP-hard.*

***Proof***: As we know that the set covering problem is NP-hard, we only prove the candidate selection problem as a set covering problem. Suppose that there is an instance of the NP-hard set covering problem: a set $U = \{u_1, u_2,..., u_l\}$, a set $CFS = \{CFS_1, CFS_2,..., CFS_k\}$ of subsets U and positive integer $d$. Is there any subset $CFS' \subseteq CFS$ of size $d$, such that each element in U belongs to at least one in $CFS'$? Next, we prove the sufficient condition and necessary condition respectively.

**Sufficient condition**: Suppose $CFS'$ is a solution to the set covering problem. We can select the corresponding set $CFS'$ of the candidates as the solution to the candidate selection problem. It is easy to obtain $u_0(CFS') = lk - |CFS'| \geq lk - d$.

**Necessary condition**: Suppose $CFS'$ is a solution to the candidate selection problem, and we have $u_0(CFS') = lk - |CFS'| \geq lk - d$. The only state that obtained this value exists when the selected candidate set covers all the candidates with the conditions $d \leq l$. So, the $CFS'$ is a solution to the set covering problem.

Hence, the theorem has been proven.

Because finding the optimal subset of candidates for data transmission in this network is NP-hard, we can exploit a heuristic algorithm to solve the problem. Next, the routing metric and the algorithm for selecting candidate set are proposed. In our scheme, we jointly consider social features, energy efficiency and ETX.

The routing metric can be described:

$$OODT_{i,j}(T) = \varphi_1 ETX_{i,j}(T) + \varphi_2 E_{iC}(T) + \varphi_3 / ST_{i,j}(T) \quad (5)$$

where $ETX_{i,j}(T)$ is the ETX of node $i$ to node $j$ in time $T$, and $\varphi_1$, $\varphi_2$, $\varphi_3 \in (0,1]$ are weight parameters, that satisfy $\varphi_1 + \varphi_2 + \varphi_3 = 1$.

In selecting candidates, the $OODT_i(T)$ of the sender and forwarding candidates must be smaller than the threshold $OODT_{threshold}(T)$ which is used as a limit to ensure a good quality link. In practice, the threshold $OODT_{threshold}(T)$ can be decided as follows:

$$OODT_{threshold}(T) = \frac{1}{|N(i)|} \sum_{j \in N(i)} OODT_{i,j}(T) \quad (6)$$

where $N(i)$ is the set of node $i$'s one-hop neighbors.

And then, we classify these selected candidates (stored in $CFS_i$) into two subsets according to OODT value, which stores the set featuring a common node belonging to different node's candidate set at the same time into $CFS_{i,2}$, and the remainder is stored in another class $CFS_{i,1}$.

The forwarder selection algorithm for any node $v$ (except for the destination node) is listed in **Algorithm 2**, where $N(v)$ is the set of $v$'s next hop nodes, $Ch(v)$ is the available channel set of node $v$ and $CFS_v$ is the forwarding candidate set of node $v$. When a sender is ready to send a packet, it inserts an extra header into this packet, which lists all nodes in $CFS$. Nodes in the $CFS$ are ranked according to their $OODT_i(T)$.

Thus, the average forwarding cost in the selected candidate set is

$$\theta_i = \frac{1}{|CFS_i|} \sum_{j \in N(i)} ST_{i,j} + ETX_i + \alpha E_{iC} \quad (7)$$

where $\alpha$ is the decay factor, which is set to be 0.01 in simulation. It is known that $E_{iC}$ is subject to the (0, 1) uniform distribution mentioned above.

Hence, we have

$$v_i = \frac{\theta_i}{\frac{1}{|CFS_i|} \sum_{j \in N(i)} ST_{i,j} + ETX_s + \alpha E_{initial}} \quad (8)$$

where $E_{initial}$ is the initial energy. $ETX_s$ is the $ETX$ of source node $s$ to the destination node. It is easy to determine that $v_i$ is subject to a (0,1) uniform distribution.

We can calculate the expected payoff of $i$ based on Bayesian Nash equilibrium solution.

$$u_i = (b_i - v_i) \prod_{j \neq i} P(b_i < (b(v_j))) \quad (9)$$

where $P(b_i < (b(v_j)))$ is the probability that $i$ has the minimum bid, and $b(v)$ is a strategy function and is assumed as a strictly increasing function of $v$.

According to the characteristics of $v_i$, we have

$$P(b_i < (b(v_j))) = P(\Phi(b_i) < v_j) = 1 - \Phi(b_i) \quad (10)$$

and

$$\prod_{j \neq i} P(b_i < (b(v_j))) = [1 - \Phi(b_i)]^{|CFS_i|-1} \quad (11)$$

---

**Algorithm 2** Forwarder Selection Algorithm (**FSA**)

---

1: $CFS_{v,1} \leftarrow \varnothing$, $CFS_{v,2} \leftarrow \varnothing$, $CFS_v \leftarrow \varnothing$
2: **for** all node $j \in N(v)$ **do**
3: calculate its OODT and the threshold according to (5), (6)
4: **if** $(OODT_i(T) \leq OODT_{threshold}(T)) \& \&(Ch(v) \cap Ch(j) \neq \varnothing) \& \&(Ch(j) \cap Ch(N(j)) \neq \varnothing)$ **then**
5: $CFS_v \leftarrow CFS_v \cup j$
6: **end if**
7: **end for**
8: divide $CFS_v$ into two subsets, the first part of $CFS_v$ stores in $CFS_{v,1}$, the latter part stores in $CFS_{v,2}$, according to the rule: whether a node belongs to the different node's candidate set at the same time or not. If yes, store it in $CFS_{v,2}$, else store it in $CFS_{v,1}$
9: **return** $CFS_{v,1}$, $CFS_{v,2}$

---

where $\Phi(b)$ is the inverse of $b(v)$.

Therefore, we have

$$\max u_i = \max_{b_i} ((b_i - v_i)[1 - \Phi(b_i)]^{|CFS_i|-1}) \quad (12)$$

Finally, we can obtain

$$b^*(v_i) = \frac{1}{|CFS_i|-1} - \frac{|CFS_i|-2}{|CFS_i|-1} v_i$$

$$= \frac{1}{|CFS_i|-1} - \frac{|CFS_i|-2}{|CFS_i|-1} \frac{\theta_i}{\frac{1}{|CFS_i|} \sum_{j \in N(i)} ST_{i,j} + ETX_s + \alpha E_{initial}} \quad (13)$$

We can see that the node $i$ bids $b^*(v_i)$ for the auction, maximizing all the bidders' payoff. Note that the node has a high priority which has a low price.

Hence, based on the routing metric and auction strategy mentioned above, the routing procedure can be described as follows:

1. When node $i$ has data to transmit, it selects some appropriate forwarders based on OODT value, and broadcasts request to them.

2. When neighbors of $i$ hear the request, they check whether the request is in the forwarder set according to **Algorithms 1** and **2**. If yes, they bid for the data transmission cost, and send back the price to node $i$. Then, the neighbors execute our proposed bid strategy. If not, the node will drop the packet. When $i$ receives the prices of the forward set, it determines the forwarding priority according to the price. Then, the node $i$ broadcasts the data packet by adding a priority list.

3. When the forwarders receive the packet, the node executes data transmission according to the forwarding strategy.

In addition, we select the best channel for data transmission according to channel availability, which can be calculated through our previous work [13]. The higher channel availability the channel has, the larger probability the channel has, which will be selected for data transmission.

## IV. PERFORMANCE EVALUATION

In this section, we compare the OODT protocol with existing works by simulation under different numbers of SUs and obstacles using the NS-2.35, CRCN model [45] and MIT reality data set [46].

The simulation settings are as follows: nodes in the network are deployed in area of 1000×1000 $m^2$. The mobile model of SUs is random waypoint model. There are 10 PUs in the network. The transmission range of the static PUs is set to 300m and their activity is modeled as a two stage ON/OFF process in which the parameter $\lambda_{busy}$ is 10. The transmission range of the SUs is set to 120m. $\chi$ is set to 0.5. The transmission standard is the IEEE 802.11b for each of the 10 channels. Log-normal shadowing is generated with standard deviation 6 dB and m=1 is used for the small-scale fading (Rayleigh) [37, 38]. The path-loss exponent parameter is 4 [47]. The SU's max speed is 2m/s. The size of the CBR data packets is 512 bytes and the destination node selected randomly. The number of radios is 2. The channel sensing time is 5ms, and the switch time is 70 $\mu s$. The bandwidth is 2 Mb/s. The duration of each run is 1000s, and the weight factors in routing metric have been set to 0.4, 0.3 and 0.3. We execute100 runs and compute the average value for performance metric in our experiments.

In addition, the parameters for energy consumption model $E_{iF}(T)$, $E_{iR}(T)$ and $E_{iACK}(T)$ have been set to 3.6e-3eu, 1.8e-3eu, and 0.16e-3eu. The initial energy of a node is 300 eu. We compare the following four protocols with the proposed OR in terms of average end-to-end delay, expected cost of routing, network lifetime and PDR: CROR [13] (coded opportunistic routing for CRAHNs), VTBR [35] (voronoi-trajectory aware routing), MDC [36] (obstacle avoiding aware routing) and SoRoute [28] (a social-aware opportunistic routing in CRAHNs).

● End-to-end delay is the time required for a packet from the source to reach the destination.

● Expected cost of routing is indicated that the expected cost of all opportunistic routing from each node to a fixed destination

● Network lifetime is the interval between the beginning of a packet transmission of the network time until the first node failure due to battery depletion

● Packet Delivery Ratio (PDR) is the ratio between the number of received packets to the number of sent packets in the network.

Next, we firstly study the impact of SU on average end-to-end delay, expected cost of routing, network lifetime and PDR. In these experiments, the number of obstacles is set to be 6.

From Fig. 7 and Fig. 8, we can see that as the number of SUs increasing, the average end-to-end delay and expected cost of routing will decrease. This is because that more SUs are used for data transmission, enhancing the data transmission. It is observed that the SoRoute has highest average end-to-end delay and expected cost of routing, which does not consider obstacle avoidance in designing OR. The OODT can achieve the smallest delay and expected cost of routing than that of others. This is because OODT exploits a novel obstacle detection algorithm and then alleviates the effect of obstacle in designing OR.

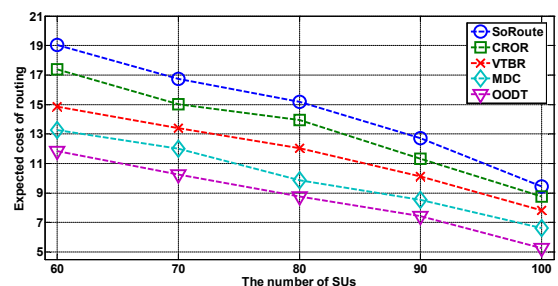

Fig. 7. Expected cost of routing vs. the number of SUs.

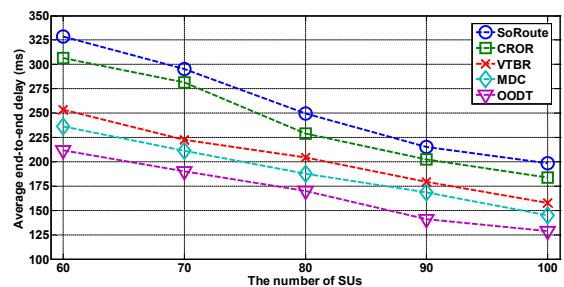

Fig. 8. Average end-to-end delay vs. the number of SUs.

In Fig. 9 and Fig. 10, we analyze the impact of the number of SUs on PDR and network lifetime. From the figures, we can see that as the number of SUs increases, the network lifetime will increase as well as shown in Fig. 10. In Fig. 9, it is observed that the PDR will increase when the number of SUs increases. For a few SUs in the network, the SU will affected by PUs, which is often isolated due to low channel availability. Hence, the PDR and network lifetime are at a low level. When more SUs appear, the routing protocols can build reliable paths for data transmission. However, the proposed scheme OODT always achieve better performance than other routing protocols, which considers obstacle avoidance and the social feature in SU's data transmission, and exploit an auction based candidate selection scheme for selecting optimal candidate to transmit data under an obstacle environment.

In Fig. 11, we study the impact of the number of obstacles on average end-to-end delay. As the number of obstacles increases, the average end-to-end delay will increase as well as shown in Fig. 11. However, the proposed OODT is better than the other schemes. This is because OODT selects proper candidates for forwarding packets based on a novel obstacle avoidance method and auction theory, which jointly considers energy, ETX and social feature in the new routing metric.

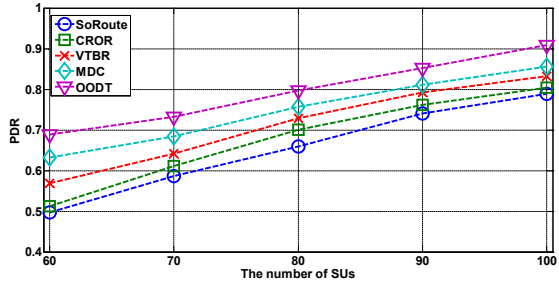

Fig. 9. PDR vs. the number of SUs.

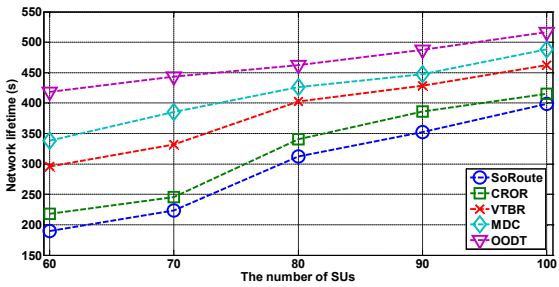

Fig. 10. Network lifetime vs. the number of SUs.

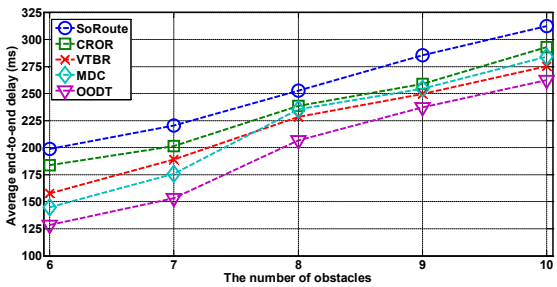

Fig. 11. Average end-to-end delay vs. the number of obstacles.

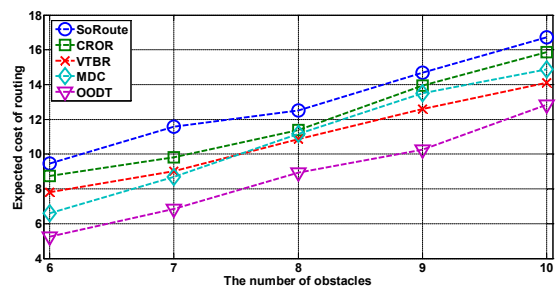

Fig. 12. Expected cost of routing vs. the number of obstacles.

Fig. 12 reflects the expected cost of routing of the five routing protocols when increasing the number of obstacles. From the figure we know that the expected cost of routing of all routing protocols increases, and the CROR, MDC, SoRoute and VTBR have the higher expected cost of routing, since they delivers data packets to all their neighbors without polygon based obstacle avoidance mechanism, hence the nodes have more chances to be selected that covered by the obstacles, and then the expected cost of routing will be larger. The OODT scheme achieves a smaller expected cost of routing than the other schemes. Because OODT exploits the polygon based obstacle avoiding method for candidate selection which considers many factors to reduce the probability that nodes' mislabeled as the candidates, and efficiently exploits auction model for candidate selection nodes in data transmission, so as to relay packets more quickly to their destinations with lower cost.

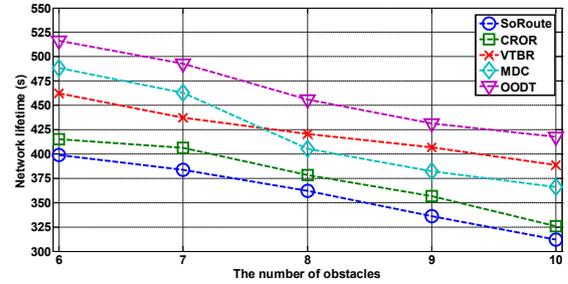

Fig. 13. Network lifetime vs. the number of obstacles.

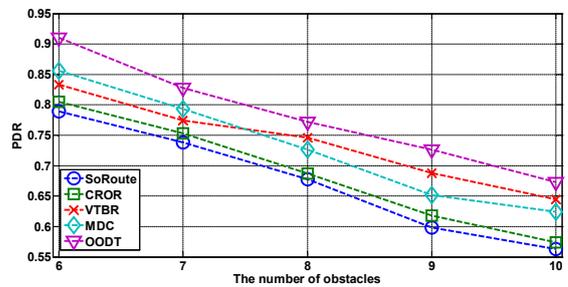

Fig. 14. PDR vs. the number of obstacles.

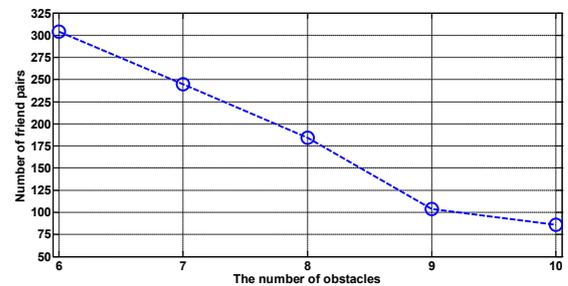

Fig.15. Number of friend pairs vs. the number of obstacles.

Fig. 13 depicts the network lifetime in these five schemes. As we can see the network time will decrease when the number of obstacles increases. However, the proposed scheme always achieves better performance that other routing protocols. This is because that they did not systematically combine obstacle avoiding and OR to improve the performance of CRAHNs for selecting path thereby show lower network lifetime.

In Fig.14, we study the impact of obstacles on PDR. As expected, we observe that the performance in terms of PDR decreases with increasing the number of obstacles. As more obstacles appear in the network, it is more difficult to find a reliable path for data transmission and then the SU's data transmission has more chances to be interrupted, thereby the PDR will decrease. In addition, the performance of VTBR will be better than MDC when the number of obstacles is larger than 8. This is because the MDC divides the whole network region into the same size grid cells, ignoring some candidate that be covered by the obstacles, while VTBR exploits a voronoi-trajectory based hybrid routing from an obstacle perspective, it will reduce the probability that mislabeled SU as a candidate. SoRoute is a social-aware opportunistic routing in CRAHNs; however, it did not consider obstacles in routing design with the lowest PDR. OODT exploits V-diagram and pruned skeletons based obstacle discovery for selecting the proper SUs as candidates. Hence, it has the highest PDR.

Fig. 15 illustrates that the number of friend pairs with obstacles. As we can see that as the number of obstacles increases, the number of friend pairs decreases. This reason is that more obstacles in the network and thus fewer nodes can transmit data by their friends. Hence, the number of friend pairs in the network will reduce.

## V. CONCLUSION

Among the surveyed existing protocols, integrating OR with the CR is an efficient strategy to improve network performance and spectrum usage. And then, we propose a novel obstacle aware OR protocol, OODT, for CRAHNs. In OODT, we exploit a novel method for characterizing candidate selection in OR, which exploits an obstacle avoiding algorithm from a computational geometry perspective. The obstacle avoiding method is based on a boundary 1-searcher's search obstacle avoiding algorithm, and its efficiency is proved. We prove that the candidate selection problem is NP-hard and then a novel auction model based candidate set selection algorithm is presented. Furthermore, the simulation results show that the OODT generally performs better than CROR, SoRoute, MDC and VTBR in terms of PDR, expected cost of routing, end-to-end delay and network lifetime. As we know in CRAHNs, the misbehavior node will attack the normal node, exaggerate the reputation of other malicious users or lower the trust level of a well-behaving node. Therefore, in future work, we will jointly consider trust and cryptography for designing OR from an energy efficient perspective.

## APPENDIX A PROOF OF THEOREM 1

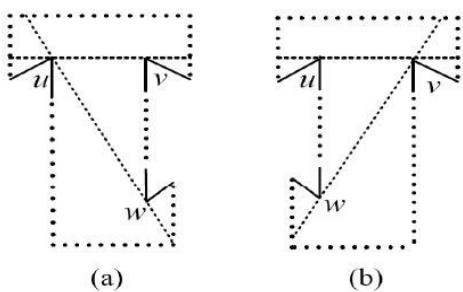

Fig 16. Two examples of the polygons satisfying C4.

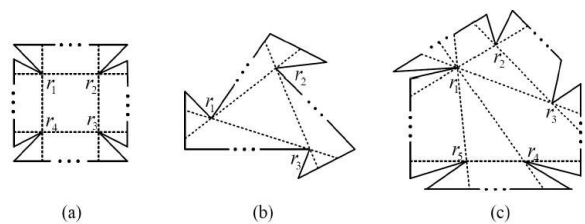

Fig 17. Examples of the polygons satisfying C1, C2 and C3 respectively.

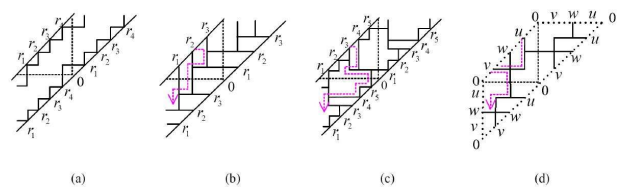

Fig 18. The pruned skeletons of Figs. 16 and 17.

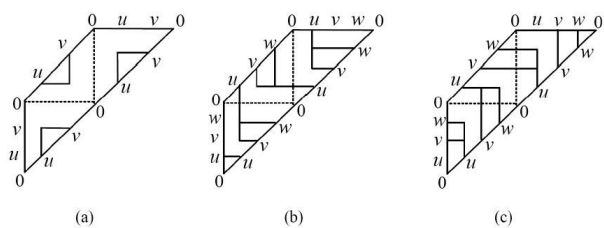

Fig 19. The corresponding pruned skeletons of bi-tangents. (a) one side bi-tangent; (b) double left bi-tangent; (c) double right bi-tangent.

**Proof: Sufficiency**: We will complete the proof of this part by considering the following four cases:

*Case 1*. $P$ satisfies $C1$. Clearly, there are at least four reflex vertices (say $r1,…, r4$) in $P$, see Figure 17(a) for an example, such that they can form at least four one-side bi-tangents and each boundary point is at least in the inner chain part of a one-side bi-tangent. Then the pruned skeleton can be expressed as shown in the Figure 18(a). From the diagram, there does not exist a search path. In general, when the number of reflex vertices in a polygon increases to $n/2$ (the maximum number of reflex vertices in $P$), and if $P$ satisfies $C1$, then its pruned skeleton is similar to Figure 18(a). Therefore, $P$ is not boundary 1-searchable by **Lemma1**.

*Case 2*. $P$ satisfies $C2$. Clearly, there are at least three reflex vertices (say $r1, r2, r3$) in $P$, see Figure 17(b) for an example, such that they can construct at least three double left bi-tangents, and each boundary point is at least in the inner chain part of a double left bi-tangent. Then the pruned skeleton can be expressed as shown in the Figure 18(b). From the diagram, there does not exist a search path. If the number of reflex vertices increases to $n/2$, and each boundary point is at least in the inner chain part of a double left bi-tangent, the pruned skeleton is similar to Figure 18(b). The situation of the double right bi-tangents is similar to that of the double left bi-tangents. Hence, $P$ is not boundary 1-searchable by **Lemma 1**.

*Case 3*. $P$ satisfies $C3$. Clearly, there are at least five reflex vertices (say $r1,…, r5$) in $P$, see Figure 17(c) for an example, where the reflex vertex pairs $(r1, r2)$, $(r2, r3)$, $(r4, r5)$, $(r5, r1)$ form four one-side bi-tangents and $r3$; $r4$ and $r1$ form a double left bi-tangent, such that some points are in the inner chain parts of one-side bi-tangents and the others are in the inner chain parts of a double left bi-tangent. Then the pruned skeleton can be expressed as shown in the Figure 18(c). From the diagram, we cannot find a search path. If the number of reflex vertices increases to $n/2$, and $P$ satisfies $C3$, then the pruned skeleton is similar to Figure 18(c). The situation of the double right bi-tangents is similar to that of the double left bi-tangents. So, $P$ is not boundary 1-searchable.

*Case 4*. $P$ satisfies $C4$. Clearly, all the polygons satisfying $C4$ are similar to the polygons shown in Figure 16(a) or 16(b). The pruned skeleton of Figure 16(a) can be expressed in Figure 18(d), and it is obvious that there is no search path. It is similar to Figure 16(b). Analogously, all the polygons satisfying $C4$ is not boundary 1-searchable.

*Necessity*: If the polygon $P$ is not boundary 1-searchable, then there is no search path in its pruned skeleton. That is, the following three cases occur in the pruned skeleton:

*Case 1*. All the points of the start line **S** of pruned skeleton are trap points. That is, every boundary point on line **S** is trapped, where a point $p$ is trapped on line **S** if any path starting from $p$ cannot be extended beyond the region without violating the *LIP*, as shown the section between $u$ and $v$ on line **S** in Figure 19(a), and the section between $v$ and $w$ on line **S** in Figure 19(c). We immediately know that the searcher should not start a search from the points of such sections. Hence, it is easy to see that if the line **S** is trapped, the searcher has nowhere to start a search. This situation implies that each boundary point is in the inner chain part of a one-side bi-tangent or a double right bi-tangent.

*Case 2*. The goal line **G** of pruned skeleton is unreachable. That is, every boundary point on line **G** is unreachable, where a point $p$ on line **G** is inside an unreachable section if any path cannot reach $p$ without violating the *LIP*, see the section between $u$ and $v$ on line **G** in Figure 19(a) and the section between $v$ and $w$ on line **G** in Figure 19(b). It is easy to see that if the line **G** is unreachable, then the searcher has nowhere to end a search. This situation implies that each boundary point is in the inner chain part of a one-side bi-tangent or a double left bi-tangent.

*Case 3*. Neither the start line **S** is trapped nor the goal line **G** is unreachable. That is, the search path from any non-trapped point on line **S** cannot reach to any point on line **G**. This implies that all the search paths starting from the non-trapped points of **S** are stopped by some bone segments, i.e., the polygon $P$ (having some double left (right) bi-tangents) satisfies $C4$. This completes the proof of **Theorem 1**.

## APPENDIX B PROOF OF THEOREM 2

To prove **Theorem 2**, we introduce the following lemmas.

**Lemma 2** [40] *A polygon $P$ is LR-visible if it is boundary 1-searchable.*

**Lemma 3** [41] *It takes $O(n)$ time to determine whether a polygon is LR-visible.*

**Lemma 4**[42] *It takes constant time to determine whether arbitrary two reflex vertices $r$ and $l$ in an LR-visible polygon such that $r \notin \partial P_{cw}[l, Back(l)], l \notin \partial P_{ccw}[r, Forw(l)]$.*

**Proof**: First, run the linear time algorithm posed in [39] to determine whether $P$ is *LR*-visible. If not, then by **Lemmas 2** and **3** it takes $O(n)$ time to determine that the polygon $P$ is not boundary 1-searchable. Otherwise, by **Theorem 1**, we just need to prove that it takes $O(n)$ time to verify whether each of $Ci$, for $i = 1, 2, 3, 4$, is established in $P$, which will be considered in the following four cases, respectively.

*Case 1*. If two visible reflex vertices $r, l$ (clockwise) form a one-side bi-tangent in an *LR*-visible polygon, then $r \notin \partial P_{cw}[l, Back(l)]$ and $l \notin \partial P_{ccw}[r, Forw(l)]$. By **Lemma 4**, we can check in constant time whether $l \notin \partial P_{ccw}[r, Forw(l)]$ and $r \notin \partial P_{cw}[l, Back(l)]$. That is, determining whether a boundary point is in the inner chain part of those bi-tangents also needs constant time. Hence, to test whether all boundary points are in the inner chain part of one-side bi-tangents needs $O(n)$ time. Therefore, it takes $O(n)$ time to verify whether $C1$ is established in $P$.

*Case 2*. First, we can compute all pairs of the disjoint backward (forward) components in $O(n)$ time [43]. Then, by the definition of left (reps. right) bi-tangent, we can compute all the pairs of reflex vertices forming left or right bi-tangents in $O(n)$ time. By the definition of the double left (resp. right) bi-tangent, we can compute all the pairs of reflex vertices forming double left or right bi-tangents in $O(n)$ time. That is, to determine whether each boundary point is in the inner chain part of a double left or right bi-tangent needs $O(n)$ time. Therefore, it takes $O(n)$ time to verify whether $C2$ is established in $P$.

*Case 3*. By the above two cases, clearly, it also takes $O(n)$ time to verify whether $C3$ is established in $P$.

*Case 4*. By Case 1, we can check in $O(n)$ time whether there exist a one-side bi-tangent constructed by $u, v$ (say) and a restricted reflex vertex $w$ (say), which is different from $u$ and $v$, in the inner chain part of this one-side bi-tangent. Then by Case 2, we can also check in $O(n)$ time whether $w$ and $u$ construct a left bi-tangent or $w$ and $v$ construct a right bi-tangent. Therefore, we can verify whether $C4$ is established in $P$ in $O(n)$ time.

In summary, we conclude that whether a polygon is boundary 1-searchable can be determined in $O(n)$ time. It is also easy to see that the space requirement is $O(n)$. This completes the proof of **Theorem 2**.

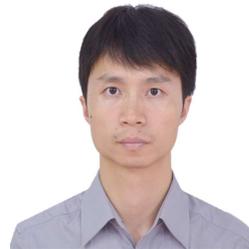

**Xiaoxiong Zhong** (S'12-M'16) received his Ph.D degree in Computer Science and Technology from Harbin Institute of Technology, China, in 2015. He was a Postdoctoral Research Fellow with Tsinghua University, from 2016 to 2018. He is currently an assistant professor with the Cyberspace Security Research Center, Peng Cheng Laboratory, Shenzhen, China. His general research interests include network protocol design and analysis, data transmission and data analysis in internet of things and edge computing, and optimization theory and algorithms for networks.

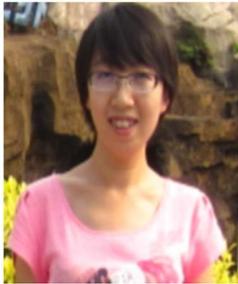 **Li Li** received her Ph.D degree in Computer Science and Technology from Harbin Institute of Technology, China, in 2017. She was a Postdoctoral Research Fellow with Tsinghua University, from 2017 to 2019. She is currently an assistant professor with the School of Computer Science, Shenzhen Institute & Information Technology, Shenzhen, China. Her research interest is in the area of wireless networks and opportunistic networks.

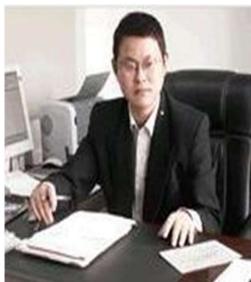 **YuanpingZhang** received his first Ph.D degree in Mathematics from Lanzhou University, China, in 1994, and the second Ph.D degree in Computer Science from The Hong Kong University of Science and Technology, (HKUST), China, in 2002. He is currently a professor with College of Computer and Educational Software, Guangzhou University, China. His research interests include algorithm design and analysis and computational geometry.

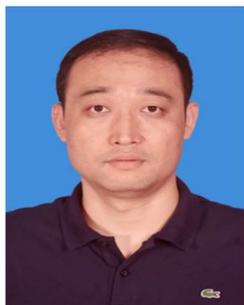 **Bin Zhang** received his Ph.D. degree in Department of Computer Science and Technology, Tsinghua University, China in 2012. He worked as a post doctor in Nanjing Telecommunication Technology Institute from 2014 to 2017. He is now a researcher in the Cyberspace Security Research Center, Peng Cheng Laboratory, Shenzhen, China. He publishes more than 30 papers in refereed international conferences and journals. His current research interests focus on network anomaly detection, Internet architecture and its protocols, information privacy security, etc.

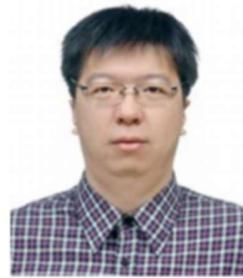 **Weizhe Zhang** (SM'18) is currently a professor in the School of Computer Science and Technology at Harbin Institute of Technology, China, and the director in the Cyberspace Security Research Center, Peng Cheng Laboratory, Shenzhen, China. His research interests are primarily in cyberspace security, cloud computing, and high-performance computing. He has published more than 160 academic papers in journals, books, and conference proceedings. He is a senior member of the IEEE and a lifetime member of the ACM.

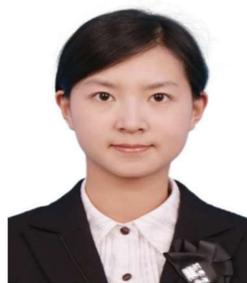 **Tingting Yang** (M'13) received the Ph.D. degrees from Dalian Maritime University, China, in 2010. She is currently a professor at Dongguan University of Technology, China. Since September 2012, she has been a visiting scholar at the Broadband Communications Research (BBCR) Lab at the Department of Electrical and Computer Engineering, University of Waterloo, Canada. Her research interests are in the areas of maritime wide band communication networks, DTN networks, and green wireless communication. She serves as the associate Editor-in-Chief of the IET Communications, as well as the advisory editor for SpringerPlus. She also serves as an associated Chair for IEEE ICC'20 & ICC'21.